\documentclass{article}
\usepackage{spconf}
\usepackage{cite}
\usepackage[pdftex]{graphicx}
\usepackage{times,amsmath}
\usepackage{color}
\usepackage{theorem}
\usepackage{amssymb}
\usepackage{subfigure}
\usepackage{hyperref}
\usepackage[ruled,vlined]{algorithm2e}
\usepackage{hyphenat}
\usepackage{tikz,chemarrow}
\usepackage[framemethod=tikz]{mdframed}
\usepackage[normalem]{ulem}

\input{my_symbol.sty}

\title{Dual-based Online Learning of Dynamic Network Topologies}
\name{Seyed Saman Saboksayr and Gonzalo Mateos\thanks{Work in this paper was supported by the NSF awards CCF-1750428 and CCF-1934962.}}
\address{Department of Electrical and Computer Engineering, University of Rochester, USA}

\begin{document}
\ninept
\maketitle
\begin{abstract}
We investigate online network topology identification from smooth nodal observations acquired in a streaming fashion. Different from non-adaptive batch solutions, our distinctive goal is to track the (possibly) dynamic adjacency matrix with affordable memory and computational costs by processing signal snapshots online. To this end, we leverage and truncate dual-based proximal gradient (DPG) iterations to solve a composite smoothness-regularized, time-varying inverse problem. Numerical tests with synthetic and real electrocorticography data showcase the effectiveness of the novel lightweight iterations when it comes to tracking slowly-varying network connectivity. We also show that the online DPG algorithm converges faster than a primal-based baseline of comparable complexity. Aligned with reproducible research practices, we share the code developed to produce all figures included in this paper.
\end{abstract}
%


\section{Introduction}
\label{S:Introduction}


The intertwined fields of graph signal processing (GSP)~\cite{ortega18, shuman13, SandryMouraSPG_TSP13}, graph representation learning~\cite{hamilton2020book}, and machine learning on graphs~\cite{chami2022machine,dong2020graph} have recently emerged with the common goal of extracting actionable information from graph-structured (i.e., relational) data describing networks~\cite[Ch. 1]{kolaczyk09}. In some real-world applications such as network neuroscience~\cite{sporns2010networks}, said relational structures may not be explicitly available~\cite{mateos19,dong2019learning}. Therefore, depending on the end goal the first step may be to recover the latent network topology to reveal patterns in the complex system under study; or, to rather learn graph representations that can facilitate downstream tasks such as classification; see e.g,~\cite{saboksayr20}. Recognizing that many of these networks are also \emph{dynamic} and that graph datasets grow every day in volume and complexity, there is a pressing need to develop efficient \emph{online} topology identification algorithms to process network data streams~\cite{giannakis18}.

The term \emph{network topology inference} encompasses a broad class of approaches to identify an underlying graph using data. Inference refers to the process of searching for a graph (represented via some graph shift operator) that is optimal for the task at hand. Data often come in the form of nodal observations (also known as graph signals in the GSP parlance), but partial edge status information is not uncommon~\cite[Ch. 7]{kolaczyk09}. Optimality notions and constraints are typically driven by statistical priors, physical laws, or explainability goals, all of which translate to models binding the observations to the sought graph. A common probabilistic prior is to model network observations via undirected Gaussian graphical models. In this case, the topology inference problem boils down to graphical model selection~\cite{dempster_cov_selec, pavez2018tsp, kumar2020jmlr}. Other recent approaches instead assume that graph signals are e.g., stationary and possibly generated by linear network diffusion~\cite{segarra2016topoidTSP16, rasoul20}, or, smooth with respect to the graph (i.e., they are sparse in the graph spectral domain)~\cite{kalofolias16, dong16, kalofolias17, kalofolias2019iclr, berger2020graphlearning}. Please refer to~\cite{mateos19, dong2019learning, giannakis18} for recent surveys of network topology inference advances.

\noindent \textbf{Proposed approach and contributions in context.} In this paper, we propose an online algorithm to track the topology of (possibly dynamic) undirected graphs using streaming, smooth signals (Section \ref{S:Preliminaries} outlines the required GSP background and formally states the topology inference problem). The rationale behind the adoption of a smoothness prior has been well-documented; see e.g.,~\cite{dong16}. Moreover, exploiting this cardinal property of network data is central to graph-based learning tasks including semi-supervised learning and denoising via variation minimization~\cite{ortega18, kolaczyk09}, just to name a few. Starting from the widely adopted, but batch, graph learning formulation in~\cite{kalofolias16, kalofolias2019iclr}, in Section \ref{S:Online} we develop a novel online dual-based proximal-gradient (DPG) algorithm that refines graph estimates sequentially-in-time. Capitalizing on favorable \emph{dual domain} structure of the smoothness-regularized inverse problem (Section \ref{S:batch})~\cite{saboksayr2021accelerated}, we leverage and truncate proximal-gradient (PG) iterations in~\cite{beck2014} to optimize the resulting time-varying cost function adaptively. Computer simulations in Section \ref{S:Simulations} showcase the effectiveness of the novel lightweight iterations when it comes to tracking slowly-varying network connectivity. The numerical experiments involve both synthetic and real electrocorticography data~\cite{kolaczyk09,kramer08}. In the interest of reproducible research, the code used to generate the figures in this paper is made publicly available. 

\noindent \textbf{Related work.} Noteworthy dynamic network topology inference algorithms exploiting a smoothness prior include~\cite{saboksayr21eusipco_ogl, cardoso20, kalofolias17, natali2022learning}. The online PG method put forth in~\cite{saboksayr21eusipco_ogl} operates in the primal domain. Yet, because of its quadratic complexity in the number of graph nodes, it is the most natural baseline to assess the proposed method's performance in tracking the optimal solution. Recently, a model-independent framework for learning time-varying graphs from online data was proposed in~\cite{natali2022learning}. Iterations therein can be accelerated by virtue of a prediction-correction strategy, but specific instances may incur cubic complexity. Unlike the novel online DPG algorithm of this paper (as well as~\cite{saboksayr21eusipco_ogl,natali2022learning}), recovery of dynamic network topology in~\cite{cardoso20, kalofolias17} is accomplished via non-recursive batch processing; hence incurring a computational cost and memory footprint that grow linearly with the number of temporal samples. For other models besides signal smoothness; see e.g.,~\cite{rasoul20,vlaski2018online,giannakis18}. 


\section{Preliminaries and Problem Statement}
\label{S:Preliminaries}


Consider a network graph $\ccalG \left ( \ccalV, \ccalE, \bbW \right)$, where $\ccalV=\{1,\ldots,N\}$ is the set of vertices and $\ccalE \subseteq \ccalV \times \ccalV$ denotes the edges. Because $\ccalG$ is assumed to be undirected, elements of $\ccalE$ are unordered pairs of vertices in $\ccalV$. The symmetric adjacency matrix $\bbW \in \reals^{N \times N}_{+}$ collects the edge weights, and $W_{ij}=0$ for $\left (i,j \right ) \notin \ccalE$. Also, $W_{ii}=0$, $\forall \:i\in\ccalV$, since we exclude self-loops. An equivalent algebraic representation of $\ccalG$'s topology is given by the graph Laplacian $\bbL := \diag \left( \bbd \right) - \bbW$, where $\bbd=\bbW\mathbf{1}$ collects the vertex degree sequence. 

\noindent\textbf{Graph signal smoothness.} We acquire graph signal observations $\bbx =\left[x_1,\dots,x_N\right]^{\top}\in\reals^N$, where $x_i$ is the value measured at $i \in \ccalV$. 
The Dirichlet energy or total variation (TV) of $\bbx$ with respect to $\bbL$ is 
\begin{equation}\label{eq5}
	\textrm{TV}(\bbx):=\bbx^{\top}\bbL \bbx
	= \frac{1}{2}\sum_{i \neq j} W_{ij} \left( x_i - x_j \right)^2.
\end{equation}
Measure $\textrm{TV}(\bbx)\in[0, \lambda_{\max}]$ quantifies the smoothness of graph signals supported on $\ccalG$~\cite{zhou04,ortega18}, where $\lambda_{\max}$ is the largest eigenvalue of the positive semidefinite matrix $\bbL$. We say a signal is smooth (or low-pass bandlimited) if it has a small total variation. From \eqref{eq5} it follows the lower bound $\textrm{TV}(\bbx)=0$ is attained by constant signals.

\noindent\textbf{Problem statement.} Given a dataset $\ccalX:=\{\bbx_t\}_{t=1}^T$ of network measurements, we want to estimate an undirected graph $\ccalG(\ccalV,\ccalE, \bbW)$ so that the signals in $\ccalX$ are smooth on $\ccalG$. We also consider tracking dynamic networks with slowly time-varying weight matrix $\bbW_t$, $t=1,2,\ldots$ ($\ccalV$ remains fixed); see Section \ref{S:Online}.


\subsection{Topology identification from smooth signals}\label{Ss:graph_learning}


We start by briefly reviewing the \emph{batch} topology inference framework in~\cite{kalofolias16, kalofolias2019iclr}, that we build on in the remainder of the paper. Consider arranging the graph signals in $\ccalX$ as columns of the data matrix $\bbX=[\bbx_1,\ldots,\bbx_T]\in \reals^{N\times T}$. Let $\bar{\bbx}_i^{\top}\in\reals^{1\times T}$ be the $i$th row of $\bbX$, which collects all $T$ observations at vertex $i$. Define the pairwise node dissimilarity matrix $\bbE\in\reals_{+}^{N\times N}$, where $E_{ij}:=\|\bar{\bbx}_i-\bar{\bbx}_j\|_2^2$, $i,j\in\ccalV$. With these definitions, it is established in~\cite{kalofolias16} that the aggregate signal smoothness measure over $\ccalX$ can be expressed as
\begin{equation}\label{E:smooth_sparse}
	\sum_{t=1}^T\textrm{TV}(\bbx_t)=\textrm{trace}(\bbX^{\top}\bbL\bbX)=\frac{1}{2}\|\bbW\circ\bbE\|_1,
\end{equation}
where $\circ$ stands for the Hadamard product. Notice how TV minimization as criterion for graph topology inference inherently induces sparsity on $\ccalE$. This is because the model preferentially selects edges $(i,j)$ with smaller pairwise nodal dissimilarities $E_{ij}$ [cf. the weighted $\ell_1$-norm in \eqref{E:smooth_sparse}]. Exploiting this intuitive relationship between signal smoothness and edge sparsity, a fairly general graph-learning framework was put forth in~\cite{kalofolias16}. The idea therein is to solve the following convex inverse problem
\begin{align}\label{eq:kalofolias}
	\min_{\bbW}&{}\:\left\{\|\bbW\circ\bbE\|_1-\alpha\bbone^{\top} \log \left( \bbW\bbone \right)+\frac{\beta}{2}\|\bbW\|_F^2\right\}\\
	\textrm{ s. to } &{} \quad\textrm{diag}(\bbW)=\mathbf{0},\: W_{ij}=W_{ji}\geq 0, \:i\neq j\nonumber
\end{align}
where $\alpha,\beta>0$ are tunable regularization parameters. In most applications, it is undesirable to have isolated (null degree) vertices in the learned graph. Hence, the logarithmic barrier imposed over the nodal degree sequence $\bbd=\bbW\bbone$. The Frobenius-norm regularization on adjacency matrix $\bbW$ offers a handle on the level of sparsity (through $\beta$). Indeed, the sparsest solution of \eqref{eq:kalofolias} will be attained by setting $\beta=0$.

Formulating \eqref{eq:kalofolias} as a search over adjacency matrices offers noteworthy computational complexity benefits. Unlike inverse problems whose optimization variable is a graph Laplacian $\bbL$~\cite{dong16}, the constraints in \eqref{eq:kalofolias} (i.e. null diagonal, symmetry and non-negativity) are all separable across the entries $W_{ij}$. This favorable structure has enabled a host of efficient \emph{batch solvers} derived based on primal-dual (PD) iterations~\cite{kalofolias16}, the PG method~\cite{saboksayr21eusipco_ogl}, and the linearized alternating-directions method of multipliers (ADMM)~\cite{wang2021}. Next, we present a dual-based algorithm we proposed in~\cite{saboksayr2021accelerated}, which comes with convergence rate guarantees in the batch setting while it is amenable to an online scheme in streaming scenarios.


\section{Proximal Gradient In The Dual Domain}\label{S:batch}


Because $\bbW$ is symmetric and has a null diagonal, the free decision variables in \eqref{eq:kalofolias} are effectively the, say, lower-triangular elements $[\bbW]_{ij}$, $j<i$. Thus, we henceforth work with the compact vector $\bbw:=\textrm{vec}[\textrm{triu}[\bbW]] \in \reals_{+}^{N(N-1)/2}$, were we have adopted convenient Matlab notation. To impose the non-negativity constraints over edge weights, we augment the cost with a penalty function
$\ind{\bbw\succeq\mathbf{0}}=0$ if $\bbw\succeq \mathbf{0}$, else $\ind{\bbw\succeq \mathbf{0}}=\infty$~\cite{kalofolias16}. Given these definitions, one can rewrite the objective in \eqref{eq:kalofolias} as the unconstrained, non-differentiable problem
\begin{equation}\label{eq:kalofolias_vec}
	\min_{\bbw}\Big\{\underbrace{\mbI\left\{\bbw\succeq\mathbf{0}\right\} + 2\bbw^{\top}\bbe+\beta\|\bbw\|_2^2}_{:=f(\bbw)}-\underbrace{\alpha \bbone^{\top} \log \left( \bbS\bbw \right)}_{ :=-g(\bbS\bbw)}\Big\}, 
\end{equation}
where $\bbe:=\textrm{vec}[\textrm{triu}[\bbE]] $ and $\bbS\in\{0,1\}^{N\times N(N-1)/2}$ maps edge weights to nodal degrees, i.e., $\bbd=\bbS\bbw$. The non-smooth function $f(\bbw):=\mbI\left\{\bbw\succeq\mathbf{0}\right\} + 2\bbw^{\top}\bbe+\beta\|\bbw\|_2^2$ is strongly convex with strong convexity parameter $2\beta$, while $g(\bbw):=-\alpha \bbone^{\top} \log \left(\bbw \right)$ is a (strictly) convex function for all $\bbw \succ \mathbf{0}$. Given the aforementioned properties of $f$ and $g$, one can establish that the composite problem \eqref{eq:kalofolias_vec} has a unique optimal solution $\bbw^\star$; see e.g.,~\cite{beck2014,wang2021}. To tackle \eqref{eq:kalofolias_vec} efficiently we can adopt a dual-based PG algorithm introduced in~\cite{beck2014}, which is capable of solving general  non-smooth, strictly convex optimization problems of the form $\min_{\bbw} \big\{f(\bbw)+g(\bbS\bbw)\big\}$. In the remainder of this section we briefly review the DPG-based graph learning framework proposed in~\cite{saboksayr2021accelerated}.

Using a standard variable-splitting technique we recast \eqref{eq:kalofolias_vec} as
\begin{equation}\label{eq:primal_split}
	\min_{\bbw,\bbd}\left\{f(\bbw)+g(\bbd)\right\},\quad  \textrm{ s. to }\bbd=\bbS\bbw.
\end{equation}
Attaching Lagrange multipliers $\bblambda\in \reals^N$ to the equality constraints and minimizing the Lagrangian function $\ccalL(\bbw,\bbd,\bblambda)=f(\bbw)+g(\bbd)-\langle\bblambda,\bbS\bbw-\bbd\rangle$ with respect to the primal variables $\{\bbw,\bbd\}$, we obtain the (minimization form) dual problem
\begin{align}
	\min_{\bblambda}&\left\{F(\bblambda)+G(\bblambda)\right\}, \text{where} \label{eq:dual} \\
	F(\bblambda):={}&\max_{\bbw}\left\{\langle \bbS^\top \bblambda,\bbw\rangle-f(\bbw)\right\}\label{eq:conjugate_F},\\
	G(\bblambda):={}&\max_{\bbd}\left\{\langle -\bblambda,\bbd\rangle-g(\bbd)\right\}.\label{eq:conjugate_G}
\end{align}
Because $f$ is strongly convex, one can derive useful smoothness properties for its Fenchel conjugate $F$. Indeed, it follows that the gradient $\nabla F(\bblambda)$ is Lipschitz continuous with constant $L:=\frac{N-1}{\beta}$; see~\cite[Lemma 1]{saboksayr2021accelerated}. Recognizing this additional structure of \eqref{eq:dual}, the PG algorithm~\cite{beck2014} (say of the ISTA type) becomes an attractive choice to solve the the dual problem. Accordingly, when applied to \eqref{eq:dual} the PG method yields the following iterations (initialized as $\bblambda_0\in \reals^N$, henceforth $k=1,2,\ldots$ denotes the iteration index)
\begin{equation}
\bblambda_{k}=\textbf{prox}_{L^{-1}G}\left(\bblambda_{k-1} - \frac{1}{L} \nabla F(\bblambda_{k-1}) \right)\label{eq:FISTA_prox},
\end{equation}
where the proximal operator of a proper, lower semi-continuous convex function $h$ is (see e.g.,~\cite{boyd14})
\begin{equation}\label{eq:prox_operator}
\textbf{prox}_{h}(\bbx)=\argmin_{\bbu}\left\{h(\bbu)+\frac{1}{2}\|\bbu-\bbx\|_2^2\right\}.
\end{equation}


\section{Online Dual Proximal Gradient Algorithm}\label{S:Online}


We switch gears to online estimation of $\bbW$ (or even tracking $\bbW_t$
in a dynamic setting) from streaming signals $\{\bbx_1,\ldots,\bbx_t,\bbx_{t+1},\ldots\}$.
A natural approach would be to solve the time-varying optimization problem at each time instant $t = 1, 2,\dots$ [cf. \eqref{eq:kalofolias_vec}]
\begin{multline}\label{eq:online}
    \bbw_{t}^{\star} \in \argmin_{\bbw} \overbrace{\left\{\mbI\left\{\bbw\succeq \mathbf{0}\right\} + 2\bbw^{\top}\bbe_{1:t} + \beta \|\bbw\|_2^2\right.}^{:=f_t(\bbw)}\\
    \underbrace{\left.-\alpha \bbone^{\top} \log \left( \bbS\bbw \right)\right\}}_{:=-g(\bbS\bbw)},
\end{multline}
where the vectorized dissimilarity matrix $\bbe_{1:t}$ is formed using all signals acquired by time $t$. As data come in, the edge-wise $\ell_1$-norm weights will fluctuate explaining the time dependence of \eqref{eq:online} through its non-smooth component $f_t$.

One can naively solve \eqref{eq:online} by sequentially running the batch DPG algorithm (cf. Section \ref{S:batch}) every time a new datum arrives. However, this approach falls short when it comes to (pseudo) real-time operation, particularly for delay-sensitive tasks that preclude running multiple $k_t=1,2,\ldots$ inner DPG iterations per time interval $[t,t+1)$ (in order to attain convergence to $\bbw_{t}^{\star}$). What is more, for dynamic networks it may not be even worth obtaining a high-precision solution $\bbw_{t}^{\star}$ (thus incurring high delay and computational burden), because at time $t+1$ a new datum arrives and the solution $\bbw_{t+1}^{\star}$ may be substantially off the prior estimate; see also Section \ref{Ss:random_graphs}. All these considerations motivate well the pursuit of an recursive algorithm that can track the solution of the time-varying optimization  \eqref{eq:online}.

Our online approach entails two steps per time instant $t=1,2,\ldots$ of data acquisition. First, we recursively update the upper-triangular entries $\bbe_{1:t}$ of the Euclidean-distance matrix once $\bbe_t$ becomes available. In stationary settings where the graph is static, it is prudent to adopt an infinite-memory scheme
\begin{equation}\label{eq_infinite_memory}
    \bbe_{1:t} = \frac{1}{t}\sum_{\tau=1}^t\bbe_\tau =\frac{\bbe_t + (t-1)\bbe_{1:t-1}}{t}. 
\end{equation}
On the other hand, in non-stationary environments arising with dynamic networks it is preferable to update $\bbe_{1:t}$ via a moving average in order to track the topology fluctuations. To that end, we employ an exponentially-weighted moving average (EWMA)
\begin{equation}\label{eq_ema}
    \bbe_{1:t} = (1 - \gamma)\bbe_{1:t-1} + \gamma \bbe_{t}, 
\end{equation}
where the constant $\gamma \in (0,1)$ is a discount (or forgetting) factor. In other words, higher $\gamma$ downweighs older observations faster. Second, we run a single iteration of the batch graph learning algorithm developed in Section \ref{S:batch} to update $\bbw_{t+1}$, namely
\begin{equation} \label{eq:onlinr_prox}
	\bblambda_{t}=\textbf{prox}_{L^{-1}G}\left(\bblambda_{t-1} - \frac{1}{L} \nabla F_t(\bblambda_{t-1}) \right),
\end{equation}
The dual variable update iteration in \eqref{eq:onlinr_prox} can be equivalently rewritten as $\bblambda_t=\bblambda_{t-1}-L^{-1}(\bbS\bbv_t-\bbu_t)$, with
\begin{align} 	
\bbv_t={}& \max\left(\mathbf{0},\frac{\bbS^\top\bblambda_{t-1}-2\bbe_{1:t}}{2\beta}\right)\label{eq:barw_update},\\ 
\bbu_t={}&\frac{\bbS\bbv_t-L\bblambda_{t-1}  + \sqrt{(\bbS\bbv_t-L\bblambda_{t-1})^2 + 4\alpha L\bbone}}{2},\label{eq:u_update}
\end{align}	
where $\max(\cdot,\cdot)$, $(\cdot)^2$, and $\sqrt{(\cdot)}$ in \eqref{eq:barw_update} are all element-wise operations of their vector arguments, see~\cite[Proposition 1]{saboksayr2021accelerated} for all details. The novel online DPG iterations are tabulated under Algorithm \ref{A:online}.

\noindent \textbf{Computational complexity.} Update \eqref{eq:barw_update} in Algorithm \ref{A:online} incurs a per iteration cost of $\ccalO(N^2)$, on par with the online PG algorithm in~\cite{saboksayr21eusipco_ogl}. The auxiliary variable $\bbu_t$ is also given in closed form, through simple operations of vectors living in the dual $N$-dimensional domain of nodal degrees [cf. the $N(N-1)/2$-dimensional primal variables $\bar{\bbw}_k$]. If additional prior knowledge about the set of possible edges is available, one can reduce the complexity further~\cite{kalofolias2019iclr}. There are no step-size parameters to tune here (on top of $\alpha$ and $\beta$) since we explicitly know the Lipschitz constant $L$~\cite[Lemma 1]{saboksayr2021accelerated}.


\begin{algorithm}[t]\label{A:online}
	\SetAlgoLined
	\textbf{Input} parameters $\alpha,\beta$, stream $\bbe_1, \bbe_2, \dots$, set $L=\frac{N-1}{\beta}$.\\
	\textbf{Initialize} $\bblambda_0$ at random. \\
	\For{$t=1,2,\dots,$}{
		Update $\bbe_{1:t}$ via either \eqref{eq_infinite_memory} or \eqref{eq_ema}\\
		$\bbv_t = \max\left(\mathbf{0},\frac{\bbS^\top\bblambda_{t-1}-2\bbe_{1:t}}{2\beta}\right)$\\
		$\bbu_t = \frac{\bbS\bbv_t-L\bblambda_{t-1}  + \sqrt{(\bbS\bbv_t-L\bblambda_{t-1})^2 + 4\alpha L\bbone}}{2}$\\
		$\bblambda_t=\bblambda_{t-1}-L^{-1}(\bbS\bbv_t-\bbu_t)$
	}
	\textbf{Output} topology estimate $\hbw_t=\max\left(\mathbf{0},\frac{\bbS^\top\bblambda_t-2\bbe_{1:t}}{2\beta}\right)$
	\caption{Online DPG for dynamic topology inference}
\end{algorithm}



\section{Numerical Experiments}\label{S:Simulations}


We perform numerical experiments to assess how well the online DPG algorithm learns random and real-world graphs, in both stationary and dynamic environments. Our main objective is to evaluate Algorithm \ref{A:online}'s effectiveness in tracking $\bbw_t^\star$ \eqref{eq:online}. This is different from explicitly monitoring the quality of the minimizer in terms of recovering the ground-truth graph used to generate the data; see e.g., the experiments in~\cite{kalofolias16, kalofolias2019iclr} for a study on the latter. This clarification notwithstanding, for the synthetic experiments we perform a rough grid search to tune $\alpha,\beta$, the criterion being to maximize the F-measure of the recovered edge set. Unless otherwise stated, the discount factor is set to $\gamma=0.002$ for all ensuing experiments. We compare Algorithm \ref{A:online} to online PG~\cite{saboksayr21eusipco_ogl}, since they both incur the same computational complexity per iteration. The code to generate all figures in this section can be downloaded from {\footnotesize{\texttt{\url{http://hajim.rochester.edu/ece/sites/gmateos/code/ODPG.zip}}}}.


\subsection{Random graphs}\label{Ss:random_graphs}


\begin{figure*}[!htb]
    \centering
    \begin{minipage}[c]{.32\textwidth}
    \includegraphics[width=\textwidth]{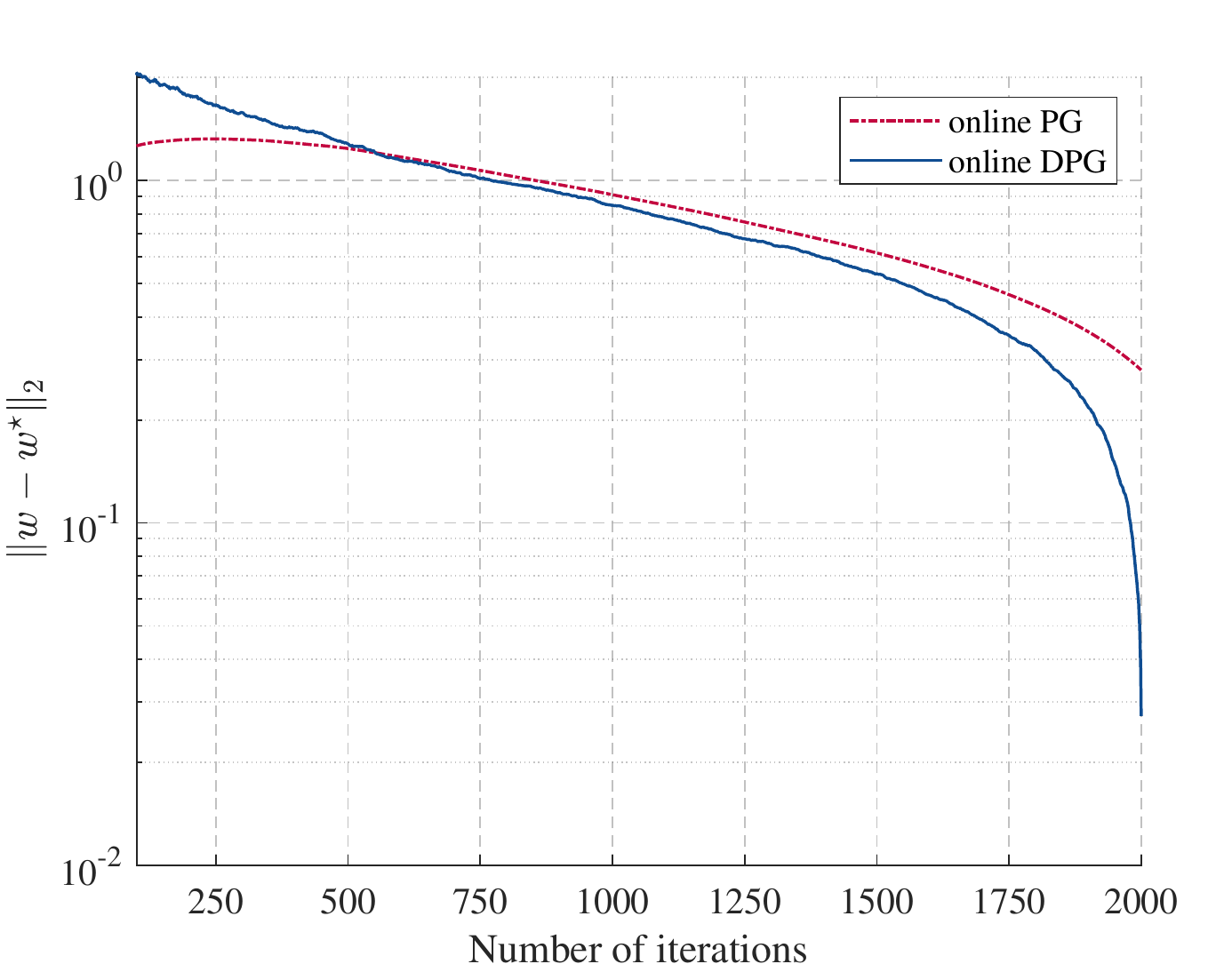}
    \newline
    \centering{\small (a)}
    \end{minipage}
    \begin{minipage}[c]{.32\textwidth}
    \includegraphics[width=\textwidth]{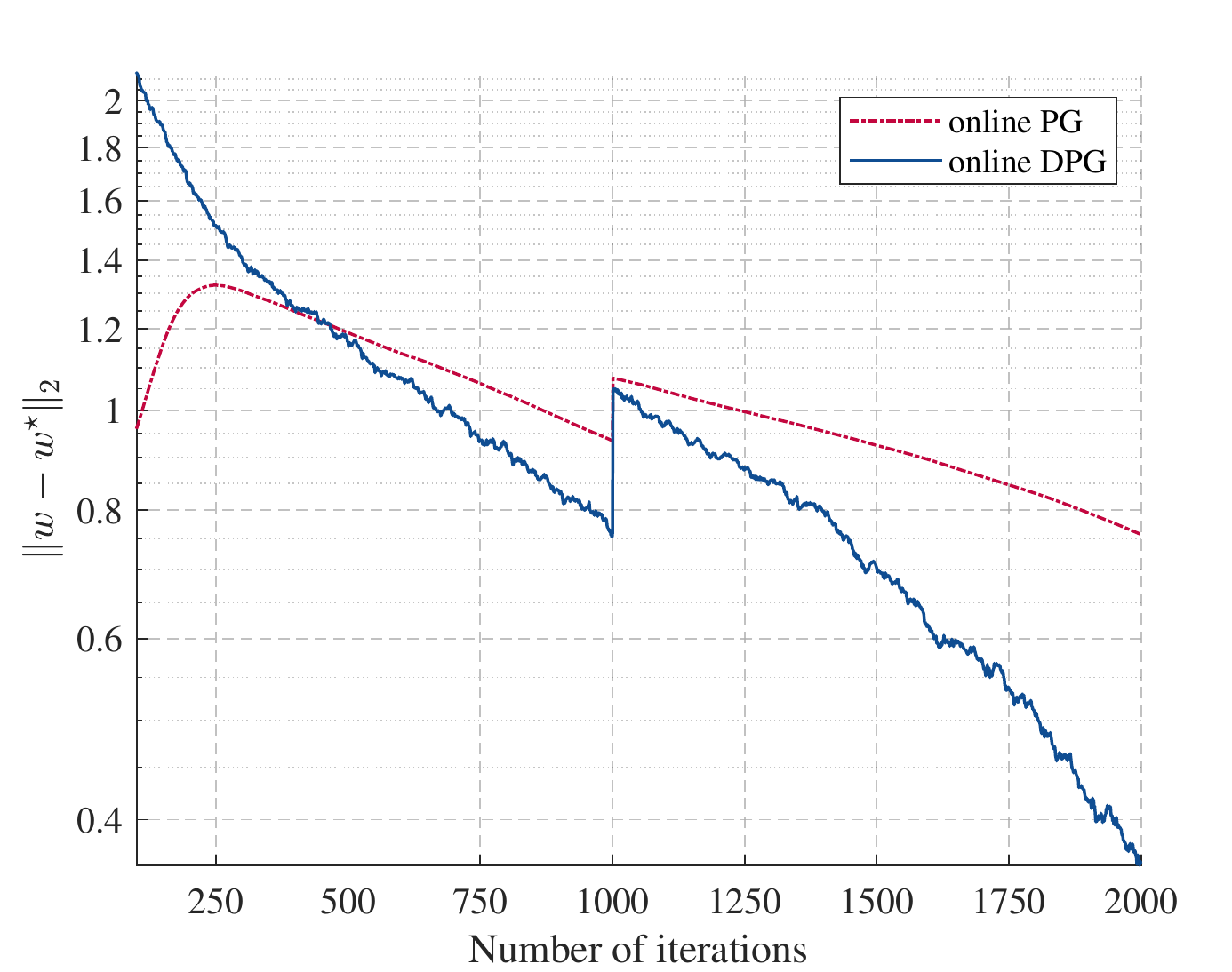}
    \newline
    \centering{\small (b)}
    \end{minipage}
    \begin{minipage}[c]{.32\textwidth}
    \includegraphics[width=\textwidth]{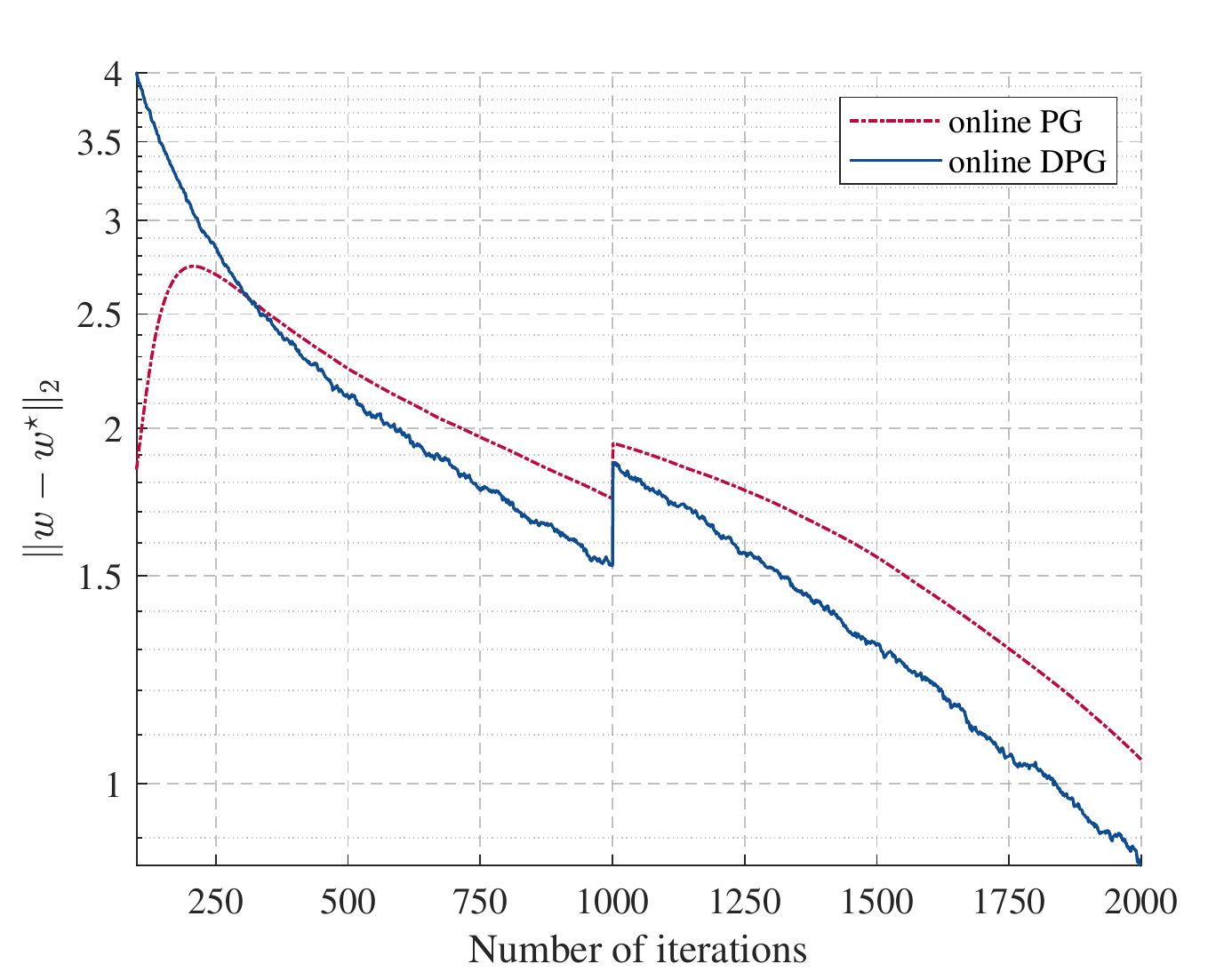}
    \newline
    \centering{\small (c)}
    \end{minipage}
    \vspace{-10pt}
    \caption{Convergence behavior illustrated via the evolution of $\|\hbw_t-\bbw_t^\star\|_2$, for various random graph models. (a) Stationary ER graph with $N=100$, (b) dynamic ER graph with $N=50$, and (c) dynamic SBM graph with $N=100$. For the dynamic network, the topology changes at $t=1000$. In all settings, the proposed online DPG method converges faster to $\bbw_t^\star$ than the baseline algorithm in~\cite{saboksayr21eusipco_ogl}.}
    \label{fig_sim}
    \vspace{-6pt}
\end{figure*}

\begin{figure*}[!htb]
    \centering
    \begin{minipage}[c]{.32\textwidth}
    \includegraphics[width=\textwidth]{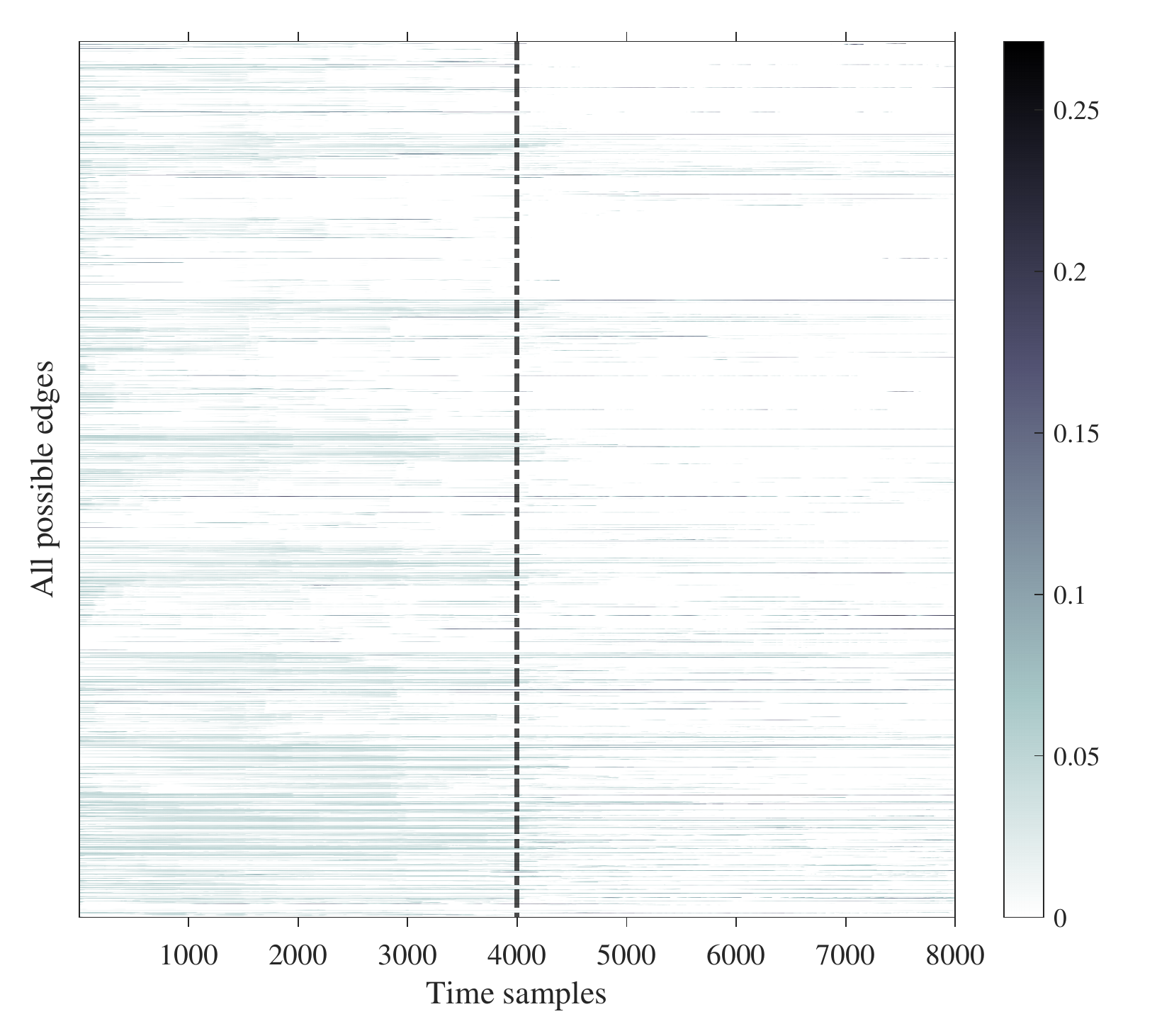}
    \newline
    \centering{\small (a)}
    \end{minipage}
    \begin{minipage}[c]{.32\textwidth}
    \includegraphics[width=\textwidth]{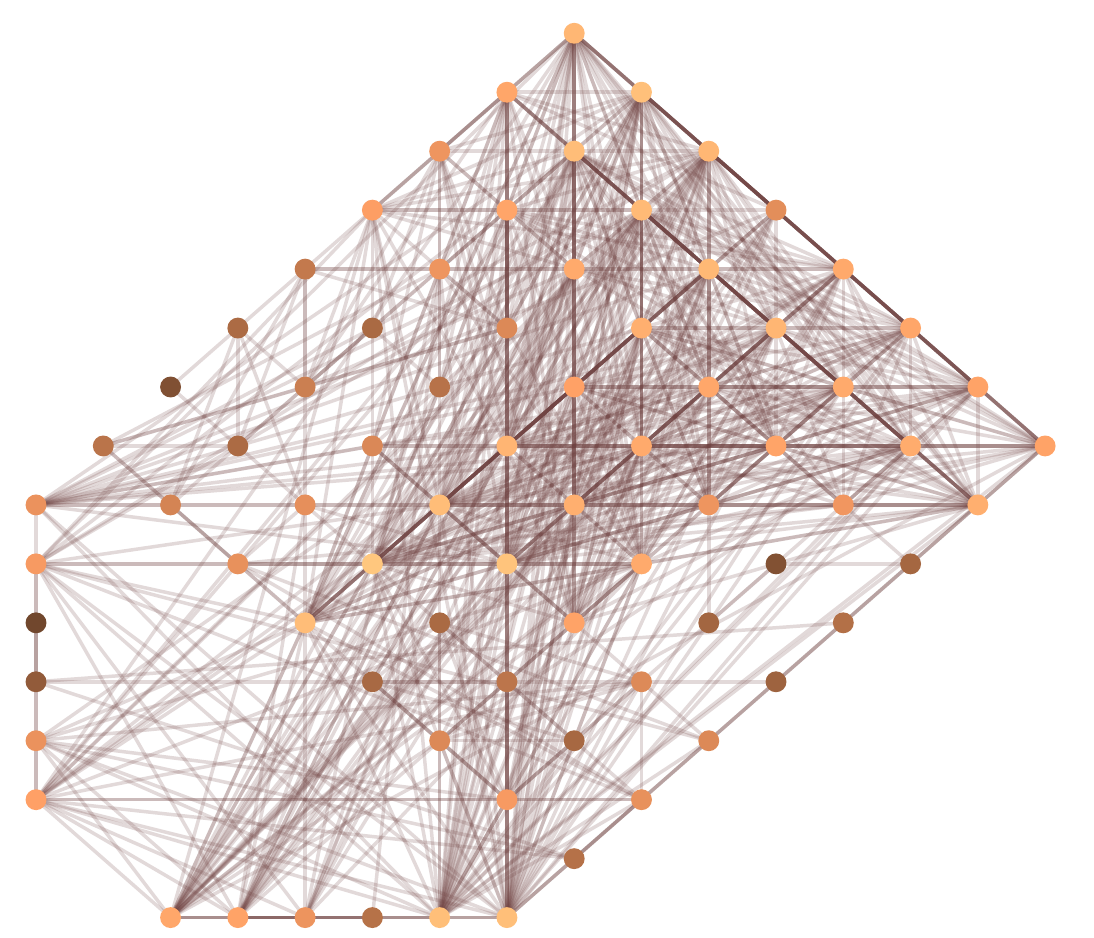}
    \newline
    \centering{\small (b)}
    \end{minipage}
    \begin{minipage}[c]{.32\textwidth}
    \includegraphics[width=\textwidth]{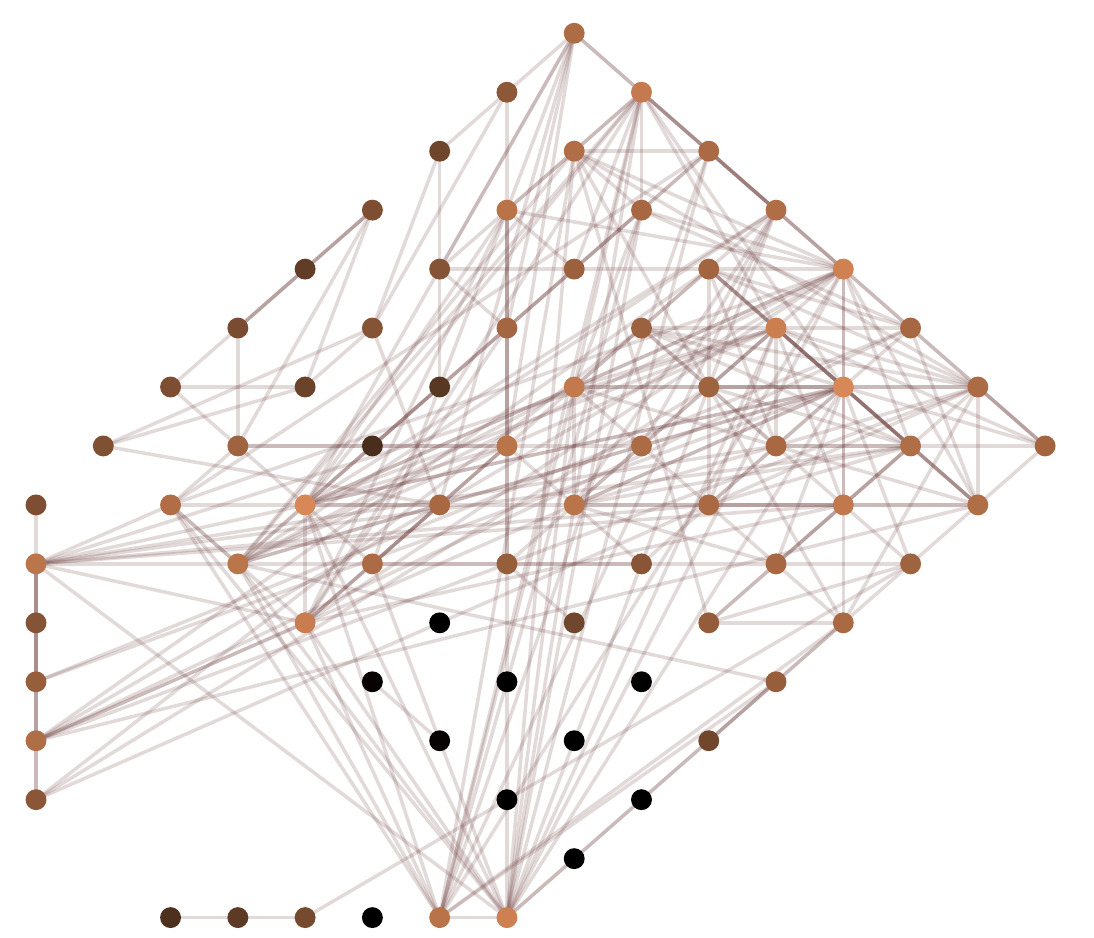}
    \newline
    \centering{\small (c)}
    \end{minipage}
    \vspace{-10pt}
    \caption{Dynamic graph learning using ECoG data. (a) Evolution of edge weights from pre-ictal to ictal stage, where the vertical line indicates seizure onset. Recovered brain graphs (b) $2.5$s prior to seizure; and (c) $2.5$s after. An edge-thinning pattern is apparent on seizure onset.}
    \label{fig_real}
    \vspace{-5pt}
\end{figure*}

Here we examine a pair of test cases. First, we assume the underlying graph is time invariant and use \eqref{eq_infinite_memory} in Algorithm \ref{A:online}. We draw a $N = 100$-node graph realization from the Erd\H{o}s-R\'enyi (ER) model (edge probability $p=0.2$). Second, we consider dynamic graph learning where the underlying network topology changes midway through the trial. We simulate two settings. A piecewise-constant sequence of two: i) random graphs is drawn from the ER model ($p=0.2$) with $N = 50$ nodes; as well as from ii) a $2$-block Stochastic Block Model (SBM) with $N=100$ vertices (even split of nodes across blocks), and connection probability $p_1=0.3$ for nodes in the same community and $p_2=0.05$ for vertices in different blocks. The initial graph switches after $t=1000$, leading to a different topology obtained after resampling $10\%$ of the edges. For $t=1,\ldots,2000$, we generate i.i.d. Gaussian-distributed smooth signals $\bbx_t \sim \ccalN\left( \bbzero, \bbL_t^{\dag}+\sigma_{e}^2 \bbI_N \right)$, where $\sigma_{e}=0.01$ and $\bbL_t$ is the Laplacian of the ground-truth graph; see e.g.,~\cite{dong16}. 

We examine the convergence behavior of the aforementioned methods by monitoring the evolution of the error metric $\| \hbw_t - \bbw_t^{\star}\|_{2}$. For given $\alpha,\beta$, we compute the solution $\bbw_t^{\star}$ by running the batch algorithm in~\cite{kalofolias16} until convergence (we use all signals available over the interval where the graph remains fixed). The results of these tests are presented Fig.~\ref{fig_sim}. The plots clearly show that Algorithm  \ref{A:online} converges faster to $\bbw_t^\star$ than the baseline~\cite{saboksayr21eusipco_ogl}; in both the stationary and dynamic settings, uniformly across model classes and graph sizes. 


\subsection{Dynamic network-based analysis of epileptic seizures}\label{Ss:real_graphs}


We test Algorithm~\ref{A:online} on an inherently dynamic task, namely a network-based study of epileptic seizures~\cite{kramer08}. To that end, we resort to a publicly available dataset that contains electrocorticogram (ECoG) signals of $8$ seizure instances acquired from human patients with epilepsy~\cite{kolaczyk09}. ECoG data are captured by $N=76$ electrodes, where $64$ of them form an $8 \times 8$ grid located at the cortical brain's surface. The other $12$ electrodes are placed deeper in the left suborbital frontal lobe and over the left hippocampal region. Signals are recorded with sampling rate of $400$ Hz. 

We apply online DPG to the ECoG signals including 10 seconds before an epileptic seizure (pre-ictal) and 10 seconds upon seizure onset (ictal). In general, we notice a significant reduction in the overall level of brain connectivity during the seizures. In Fig.~\ref{fig_real}(a) we plot the temporal variation of edge weights as we update the recovered dynamic brain network. The vertical line indicates the moment of seizure onset, where the drop in edge density is considerable. Moreover, in Figs.~\ref{fig_real}(b) and (c) we depict two snapshots of the learned graph at $2.5$ seconds prior to seizure and $2.5$ seconds after; respectively. Vertex colors in Fig.~\ref{fig_real}(b) are proxies of closeness centrality values, where darker shades correspond to lower values. The closeness centrality scores are calculated as the inverse sum of the distances from a node to all other nodes~\cite[Ch.4]{kolaczyk09}. We observe in Fig.~\ref{fig_real}(b) and (c) that edge thinning is more prominent in the bottom corner of the grid and along the two strips. This result is well aligned with the findings in~\cite{kramer08}.

\vspace{-5pt}
\section{Concluding Summary}\label{S:conclusions}


We proposed an online algorithm to track the topology of slowly-varying undirected graphs from streaming signals. Capitalizing on favorable dual domain structure of a smoothness-regularized inverse problem with well-documented merits, we derive and truncate proximal gradient iterations to minimize a time-varying cost in an online fashion. The novel algorithm is devoid of (often hard to tune) step-sizes, it is lightweight and demands constant memory storage regardless of the number of measurements. Numerical tests with synthetic and real brain activity signals demonstrate the effectiveness of the dynamic network topology tracker, and that it compares favorably against a state-of-the-art baseline of comparable complexity.


\newpage

\bibliographystyle{IEEEtran}
%
\bibliography{refs}

\begin{thebibliography}{10}
\providecommand{\url}[1]{#1}
\csname url@samestyle\endcsname
\providecommand{\newblock}{\relax}
\providecommand{\bibinfo}[2]{#2}
\providecommand{\BIBentrySTDinterwordspacing}{\spaceskip=0pt\relax}
\providecommand{\BIBentryALTinterwordstretchfactor}{4}
\providecommand{\BIBentryALTinterwordspacing}{\spaceskip=\fontdimen2\font plus
\BIBentryALTinterwordstretchfactor\fontdimen3\font minus
  \fontdimen4\font\relax}
\providecommand{\BIBforeignlanguage}[2]{{%
\expandafter\ifx\csname l@#1\endcsname\relax
\typeout{** WARNING: IEEEtran.bst: No hyphenation pattern has been}%
\typeout{** loaded for the language `#1'. Using the pattern for}%
\typeout{** the default language instead.}%
\else
\language=\csname l@#1\endcsname
\fi
#2}}
\providecommand{\BIBdecl}{\relax}
\BIBdecl

\bibitem{ortega18}
A.~{Ortega}, P.~{Frossard}, J.~Kova\u{c}evi\'{c}, J.~M.~F. {Moura}, and
  P.~{Vandergheynst}, ``Graph signal processing: Overview, challenges, and
  applications,'' \emph{Proc. IEEE}, vol. 106, no.~5, pp. 808--828, 2018.

\bibitem{shuman13}
D.~I. {Shuman}, S.~K. {Narang}, P.~{Frossard}, A.~{Ortega}, and
  P.~{Vandergheynst}, ``The emerging field of signal processing on graphs:
  Extending high-dimensional data analysis to networks and other irregular
  domains,'' \emph{IEEE Signal Process. Mag.}, vol.~30, no.~3, pp. 83--98,
  2013.

\bibitem{SandryMouraSPG_TSP13}
A.~Sandryhaila and J.~M.~F. Moura, ``Discrete signal processing on graphs,''
  \emph{IEEE Trans. Signal Process.}, vol.~61, no.~7, pp. 1644--1656, Apr.
  2013.

\bibitem{hamilton2020book}
W.~L. Hamilton, ``Graph representation learning,'' \emph{Synth. Lect. Artif.
  Intell. Mach. Learn.}, vol.~14, no.~3, pp. 1--159, 2020.

\bibitem{chami2022machine}
I.~Chami, S.~Abu-El-Haija, B.~Perozzi, C.~R{\'e}, and K.~Murphy, ``Machine
  learning on graphs: A model and comprehensive taxonomy,'' \emph{J. Mach.
  Learn. Res.}, vol.~23, no.~89, pp. 1--64, 2022.

\bibitem{dong2020graph}
X.~Dong, D.~Thanou, L.~Toni, M.~Bronstein, and P.~Frossard, ``Graph signal
  processing for machine learning: A review and new perspectives,'' \emph{IEEE
  Signal Processing Magazine}, vol.~37, no.~6, pp. 117--127, 2020.

\bibitem{kolaczyk09}
E.~D. Kolaczyk, \emph{Statistical Analysis of Network Data: Methods and
  Models}.\hskip 1em plus 0.5em minus 0.4em\relax New York, NY:
  Springer\hyp{}Verlag, 2009.

\bibitem{sporns2010networks}
O.~Sporns, \emph{Networks of the Brain}.\hskip 1em plus 0.5em minus 0.4em\relax
  MIT Press, 2010.

\bibitem{mateos19}
G.~{Mateos}, S.~{Segarra}, A.~G. {Marques}, and A.~{Ribeiro}, ``Connecting the
  dots: Identifying network structure via graph signal processing,'' \emph{IEEE
  Signal Process. Mag.}, vol.~36, no.~3, pp. 16--43, 2019.

\bibitem{dong2019learning}
X.~{Dong}, D.~{Thanou}, M.~{Rabbat}, and P.~{Frossard}, ``Learning graphs from
  data: A signal representation perspective,'' \emph{IEEE Signal Process.
  Mag.}, vol.~36, no.~3, pp. 44--63, 2019.

\bibitem{saboksayr20}
S.~S. Saboksayr, G.~Mateos, and M.~Cetin, ``Online discriminative graph
  learning from multi-class smooth signals,'' \emph{Signal Process.}, vol. 186,
  p. 108101, 2021.

\bibitem{giannakis18}
G.~B. Giannakis, Y.~Shen, and G.~V. Karanikolas, ``Topology identification and
  learning over graphs: Accounting for nonlinearities and dynamics,''
  \emph{Proc. IEEE}, vol. 106, no.~5, pp. 787--807, 2018.

\bibitem{dempster_cov_selec}
A.~P. Dempster, ``Covariance selection,'' \emph{Biometrics}, vol.~28, no.~1,
  pp. 157--175, 1972.

\bibitem{pavez2018tsp}
E.~Pavez, H.~E. Egilmez, and A.~Ortega, ``Learning graphs with monotone
  topology properties and multiple connected components,'' \emph{IEEE Trans.
  Signal Process.}, vol.~66, no.~9, pp. 2399--2413, May 2018.

\bibitem{kumar2020jmlr}
S.~Kumar, J.~Ying, J.~V. de~M.~Cardoso, and D.~P. Palomar, ``A unified
  framework for structured graph learning via spectral constraints,'' \emph{J.
  Mach. Learn. Res.}, vol.~21, no.~22, pp. 1--60, 2020.

\bibitem{segarra2016topoidTSP16}
S.~Segarra, A.~Marques, G.~Mateos, and A.~Ribeiro, ``Network topology inference
  from spectral templates,'' \emph{IEEE Trans. Signal Inf. Process. Netw.},
  vol.~3, no.~3, pp. 467--483, Aug. 2017.

\bibitem{rasoul20}
R.~Shafipour and G.~Mateos, ``Online topology inference from streaming
  stationary graph signals with partial connectivity information,''
  \emph{Algorithms}, vol.~13, no.~9, pp. 1--19, Sep. 2020.

\bibitem{kalofolias16}
V.~Kalofolias, ``How to learn a graph from smooth signals,'' in \emph{Artif.
  Intel. and Stat. (AISTATS)}, 2016, pp. 920--929.

\bibitem{dong16}
X.~{Dong}, D.~{Thanou}, P.~{Frossard}, and P.~{Vandergheynst}, ``Learning
  {L}aplacian matrix in smooth graph signal representations,'' \emph{IEEE
  Trans. Signal Process.}, vol.~64, no.~23, pp. 6160--6173, 2016.

\bibitem{kalofolias17}
V.~{Kalofolias}, A.~{Loukas}, D.~{Thanou}, and P.~{Frossard}, ``Learning time
  varying graphs,'' in \emph{IEEE Intl. Conf. Acoust., Speech and Signal
  Process. (ICASSP)}, 2017, pp. 2826--2830.

\bibitem{kalofolias2019iclr}
V.~Kalofolias and N.~Perraudin, ``Large scale graph learning from smooth
  signals,'' in \emph{Int. Conf. Learning Representations (ICLR)}, 2019.

\bibitem{berger2020graphlearning}
P.~{Berger}, G.~{Hannak}, and G.~{Matz}, ``Efficient graph learning from noisy
  and incomplete data,'' \emph{IEEE Trans. Signal Inf. Process. Netw.}, vol.~6,
  pp. 105--119, 2020.

\bibitem{saboksayr2021accelerated}
S.~S. Saboksayr and G.~Mateos, ``Accelerated graph learning from smooth
  signals,'' \emph{IEEE Signal Process. Lett.}, vol.~28, pp. 2192--2196, 2021.

\bibitem{beck2014}
A.~Beck and M.~Teboulle, ``A fast dual proximal gradient algorithm for convex
  minimization and applications,'' \emph{Operations Research Letters}, vol.~42,
  no.~1, pp. 1--6, 2014.

\bibitem{kramer08}
M.~A. Kramer, E.~D. Kolaczyk, and H.~E. Kirsch, ``Emergent network topology at
  seizure onset in humans,'' \emph{Epilepsy Res.}, vol.~79, no.~2, pp.
  173--186, 2008.

\bibitem{saboksayr21eusipco_ogl}
S.~S. Saboksayr, G.~Mateos, and M.~Cetin, ``Online graph learning under
  smoothness priors,'' in \emph{European Signal Process. Conf. (EUSIPCO)},
  Dublin, Ireland, 2021, pp. 1820--1824.

\bibitem{cardoso20}
J.~V. d.~M. Cardoso and D.~P. Palomar, ``Learning undirected graphs in
  financial markets,'' in \emph{Asilomar Conf. Signals, Systems, and
  Computers}, 2020, pp. 741--745.

\bibitem{natali2022learning}
A.~Natali, E.~Isufi, M.~Coutino, and G.~Leus, ``Learning time-varying graphs
  from online data,'' \emph{IEEE Open J. Signal Process.}, vol.~3, pp.
  212--228, 2022.

\bibitem{vlaski2018online}
S.~{Vlaski}, H.~P. {Mareti{\'c}}, R.~{Nassif}, P.~{Frossard}, and A.~H.
  {Sayed}, ``Online graph learning from sequential data,'' in \emph{IEEE Data
  Sci. Wrksp. (DSW)}, 2018.

\bibitem{zhou04}
D.~Zhou and B.~Sch{\"o}lkopf, ``A regularization framework for learning from
  graph data,'' in \emph{Int. Conf. Mach. Learning (ICML)}, 2004.

\bibitem{wang2021}
X.~Wang, C.~Yao, H.~Lei, and A.~M.-C. So, ``An efficient alternating direction
  method for graph learning from smooth signals,'' in \emph{IEEE Intl. Conf.
  Acoust., Speech and Signal Process. (ICASSP)}, Toronto, Canada, 2021, pp.
  5380--5384.

\bibitem{boyd14}
N.~Parikh and S.~Boyd, ``Proximal algorithms,'' \emph{Foundations and Trends in
  optimization}, vol.~1, no.~3, p. 127–239, 2014.

\end{thebibliography}

\end{document}